\documentclass[twocolumn,aps,prl,superscriptaddress,reprint,longbibliography]{revtex4-1}
\usepackage{soul}
\usepackage{afterpage}

\usepackage{float}
\usepackage[breaklinks,colorlinks,allcolors=blue]{hyperref}
\usepackage{placeins}
\usepackage{dsfont}
\usepackage[utf8]{inputenc}
\usepackage{bm}
\usepackage{amssymb}
\usepackage{amsmath}
\usepackage{amsfonts}
\usepackage{graphicx}
\usepackage[usenames,dvipsnames]{xcolor} 
\usepackage{dcolumn}
\usepackage{bm}

\newcommand{\tr}[1]{\text{Tr}[ #1]}
\newcommand{\ket}[1]{| #1 \rangle}

\newcommand{\braket}[1]{ \langle{{#1}}\rangle}

\newcommand{\op}[1]{\mathcal{ #1 }}
\newcommand{\Id}{1\!\!1}

\usepackage[normalem]{ulem}

\begin{document}
\title{Dissipation-assisted operator evolution method for capturing hydrodynamic transport}
\makeatletter
\let\inserttitle\@title
\makeatother

\author{Tibor Rakovszky}

\affiliation{Department of Physics, T42, Technische Universit{\"a}t M{\"u}nchen, James-Franck-Stra{\ss}e 1, D-85748 Garching, Germany}

\author{C.W.~von~Keyserlingk}

\affiliation{School of Physics \& Astronomy, University of Birmingham, Birmingham, B15 2TT,
UK}

\author{Frank Pollmann}

\affiliation{Department of Physics, T42, Technische Universit{\"a}t M{\"u}nchen, James-Franck-Stra{\ss}e 1, D-85748 Garching, Germany}
\affiliation{Munich Center for Quantum Science and Technology (MCQST), Schellingstr. 4, D-80799 M\"unchen, Germany }

\begin{abstract}

We introduce the dissipation-assisted operator evolution (DAOE) method for calculating transport properties of strongly interacting lattice systems in the high temperature regime. DAOE is based on evolving observables in the Heisenberg picture, and applying an artificial dissipation that reduces the weight on non-local operators. We represent the observable as a matrix product operator, and show that the dissipation leads to a decay of operator entanglement, allowing us to capture the dynamics to long times. We test this scheme by calculating spin and energy diffusion constants in a variety of physical models. By gradually weakening the dissipation, we are able to consistently extrapolate our results to the case of zero dissipation, thus estimating the physical diffusion constant with high precision.

\end{abstract}

\maketitle

\paragraph{Introduction.---} Despite their complexity, thermalizing quantum many-body systems often exhibit universal hydrodynamical features in their low-frequency, long-wavelength limit~\cite{ChaikinLubensky,ForsterBook,Kadanoff1963,Bertini2020Review,LiuNotes,Delacretaz2019,DoyonNotes,Lux2013}. Although these features are routinely measured in transport experiments, quantitatively connecting them to the underlying microscopic dynamics, e.g., deriving the transport coefficients from first principles in generic interacting quantum systems, is notoriously difficult in practice~\cite{ForsterBook,Pomeau1975,Rau1996,Zwanzig2001,Grabert1982,Banks2019}. It usually requires evaluating dynamical correlations, and existing methods are typically restricted to small systems or short times, often leading to unreliable results~\cite{Bertini2020Review,Heitmann2020,Schollwock2011,Paeckel2019,Viswanath1994}. While methods have been proposed to circumvent these issues in certain cases~\cite{Prosen2009,Znidaric2019,Haegeman2011,Haegeman2016,Leviatan2017,Kloss2018,White2018,Wurtz2018,Krumnow2019,Parker2019}, calculating transport properties reliably in generic models remains a challenging task.

The purpose of this paper is to introduce a numerical method for calculating transport properties from first principles in a controlled manner, while avoiding finite size and time restrictions. We achieve this by focusing on the Heisenberg picture dynamics of conserved densities. Motivated by recent results on operator spreading~\cite{Nahum2017,RvK2017,Khemani2018,RvK2018}, we introduce an artificial dissipation that removes operators based on their spatial support. As a consequence, the time-evolved operator may be stored more compactly using standard tensor network techniques. The resulting dynamics depends on the specifics of the dissipative procedure, but in the limit of weak dissipation, the different methods all appear to converge. This allows us to estimate the physical result (here, a spin or energy diffusion constant) through extrapolation.

\begin{figure}
\centering
	\includegraphics[width=1.\columnwidth]{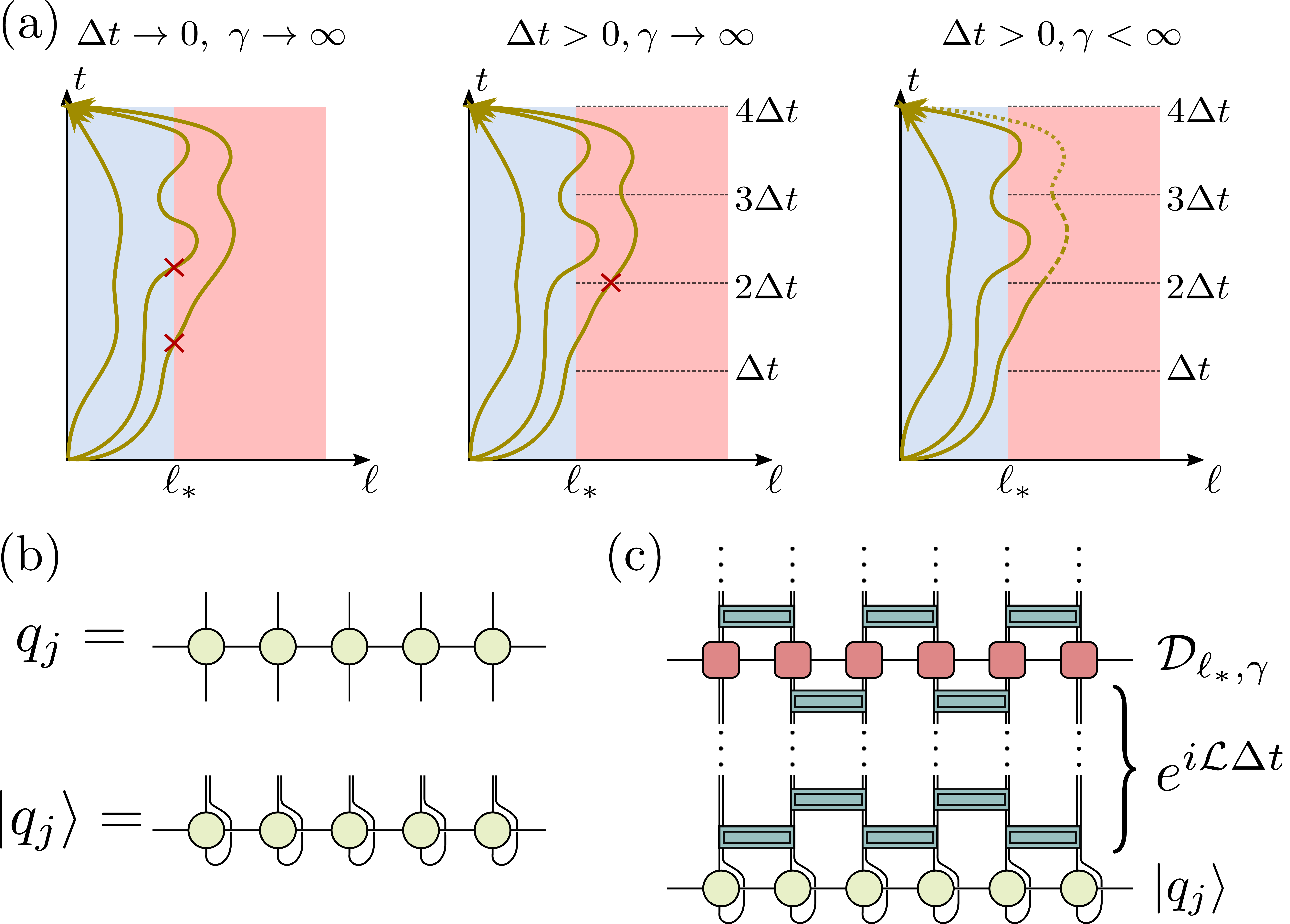}
	\caption{Dissipation-assisted operator evolution (DAOE) method. (a) Sketch of the non-unitary evolution~\eqref{eq:DissipativeEvol} as a sum over paths in operator space. For $\Delta t \to 0$, $\gamma \to \infty$, paths that leave the $\ell \leq \ell_*$ subspace are discarded immediately. Making $\Delta t$ finite, we keep paths that wander off from this subspace but return before the next integer multiple of $\Delta t$. Finally, when $\Delta t, \gamma$ are both finite, all paths are kept, but the weight of those that spend time outside the `slow' subspace is gradually reduced. (b): The operator (MPO) $q_j$ can be reinterpreted as a state (MPS) $\ket{q_j}$ on a doubled Hilbert space. (c): One period of the DAOE as a tensor network. $\ket{q_j}$ is evolved with the TEBD algorithm up to time $\Delta t$. Then the dissipator $\op{D}_{\ell_*,\gamma}$ is applied as a bond dimension $\ell_*+1$ MPO.}
	\label{fig:Method}
\end{figure}

\paragraph{Numerical method.---} We work with one-dimensional lattice models, labeling sites by $j=1,\ldots,L$. Consider the local density, $q_j$, of some conserved quantity $Q = \sum_j q_j$ (e.g., charge or energy). We are interested in dynamical correlations of these densities, $\braket{q_i(0) q_j(t)}_\text{eq}$, evaluated in some equilibrium state. We focus on infinite temperature, so that $\braket{\ldots}_\text{eq} \equiv \tr{\ldots} / \mathcal{N}$, with $\mathcal{N}$ the Hilbert space dimension. Here $q_j(t)$ is evolved unitarily in the Heisenberg picture, with a Hamiltonian $H$ that conserves $Q$, $[H,Q]=0$. Transport properties can be extracted from such correlations, as we detail below. 

In what follows, we shall find it useful to think of operators as vectors in an enlarged Hilbert space of size $\mathcal{N}^2$. In a matrix product operator (MPO) representation, this is equivalent to combining the two physical legs into a single leg, turning it into a matrix product \emph{state} (MPS), as illustrated by Fig.~\ref{fig:Method}(b). We use the notation $\ket{q_j}$ for the `vectorized' operator, and introduce an inner product on this space as $\braket{A|B} \equiv \braket{A^\dagger B}_\text{eq}$. The Heisenberg equation of motion can be rewritten as $\partial_t \ket{q_j} = i [H,q_j] \equiv i \op{L} \ket{q_j}$, which defines the \emph{Liouvillian} superoperator, $\op{L}$. This is solved by $\ket{q_j(t)} = e^{i\op{L}t} \ket{q_j}$. The key point is that we do not need the full evolution $e^{i\op{L}t}$ -- we are only interested in its matrix elements in the `slow' subspace, spanned by the conserved densities: $\braket{q_i | e^{i\op{L}t} | q_j} = \braket{q_i q_j(t)}_\text{eq}$. This \emph{projected} evolution is generically no longer unitary.

We wish to approximate this non-unitary evolution by gradually taking into account the effect of the `bath', meaning all the remaining operators that we are not projecting onto. We will do this in a more general way, where we include not only conserved densities, but all sufficiently local operators in the slow subspace. To be concrete, let us imagine a spin-$1/2$ chain. Then a basis of all $4^{L}$ operators is given by \emph{Pauli strings}, products of the four Pauli matrices $\Id , X, Y, Z$. To each Pauli string $\op{S}$, we can associate a \emph{length} $\ell_{\op{S}}$, which is simply the number of non-trivial Paulis occurring in it. For example, $1\!\!1, X_j, Z_iY_j$ have lengths $\ell=0,1,2$, respectively. We can then define a \emph{dissipation superoperator} that decreases the weight of all strings longer than some cutoff length $\ell_*$ as
\begin{align}
\op{D}_{\ell_*,\gamma}[\op{S}] = 
\begin{cases}
\op{S} &\text{ if } \ell_\op{S} \leq \ell_* \\ 
e^{-\gamma(\ell_\op{S} - \ell_*)} \op{S} &\text{ otherwise}.
\end{cases}
\end{align}
The cutoff length $\ell_*$ is introduced to ensure that the physically most relevant operators, such as conserved densities, are not affected by the dissipator.

We are now in a position to describe our proposed method. We define a modified time evolution, by applying the dissipator with period $\Delta t$. That is, for time $t \in[ N, N+1) \Delta t$ (for $N \in \mathbb{N}$), we consider the time evolved local density defined by 
\begin{equation}\label{eq:DissipativeEvol}
\ket{\tilde q_j(t)} = e^{i\mathcal{L} (t-N\Delta t)} \left( \op{D}_{\ell_*,\gamma}  e^{i\mathcal{L} \Delta t}\right)^N \ket{q_j};
\end{equation}
we dub this \emph{dissipation-assisted operator evolution} (DAOE). Eq.~\eqref{eq:DissipativeEvol} is clearly very different from the true, unitarily evolved operator $\ket{q_j(t)}$. However, we propose that the dissipative evolution can be made to correctly capture the correlations with other slow operators, particularly conserved densities, $\braket{q_i|\tilde q_j(t)} \approx \braket{q_i | q_j(t)}$.

Intuitively, $\Delta t$ and $1/\gamma$ both play a similar role, limiting the amount of time an operator is allowed to spend outside the $\ell \leq \ell^*$ subspace. While at $\Delta t \to 0$, $\gamma \to \infty$ the dynamics is projected down to this subspace~\footnote{Approximations of this sort have been used to study magnetic resonance~\cite{Kuprov2007,Karabanov2011}}, making either $\Delta t$ or $\gamma$ finite allows the operators to go outside, but only for a limited amount of time (in fact, when $\gamma$ is small, results depend on the ratio $\gamma/\Delta t$ only~\cite{suppmat}). One can think of this as summing up certain contributions in a path-integral representation of the propagator $\braket{q_i | e^{i\op{L}t} | q_j}$, as illustrated in Fig.~\ref{fig:Method}(a).

Unitary evolution is recovered by taking either $\gamma \to 0$, $\Delta t \to \infty$ or $\ell_* \to \infty$. In practice, we shall find it most useful to take the first option, keeping $\ell_*$ and $\Delta t$ fixed while approaching the unitary limit through decreasing $\gamma$. We expect DAOE to achieve a good approximation to $\braket{q_i | q_j(t)}$ for the following reasons. The correlators considered are affected by the dissipation through `backflow' processes~\cite{Khemani2018}, wherein a long Pauli string in $q_j(t')$ at time $t' < t$ develops a component on a short operator such as $q_i$ by time $t$. Random circuit calculations~\cite{Khemani2018,RvK2018} suggest that the combination of entropic effects and dephasing makes the backflow contributions to the diffusion constant decay exponentially with $\ell_*$. We leave a detailed discussion of this to future work~\cite{RvK_future}.

The spirit of our approximation is closely related to the well-known \emph{memory matrix formalism} (MMF)~\cite{Zwanzig1961,Mori1965,ForsterBook,Zwanzig2001,Pomeau1975,Grabert1982,Rau1996}. The `short' ($\ell \leq \ell_*$) and `long' ($\ell > \ell_*$) operators play the role of the `slow' and `fast' subspaces in the MMF. The backflow processes mentioned above, where a short operator grows to a long string and then back, are then similar to the kind of memory effects that MMF delineates. The dissipation acts as a cutoff on the memory time, more strongly suppressing processes where the operator is 'long' for a greater duration. This is expected to be a good approximation if these processes are indeed `fast'. However, our method is directly applicable to generic lattice models in 1D, unlike the MMF, where one either perturbs around some fine-tuned point~\cite{Jung2006,Jung2007}, or makes some uncontrolled approximations to simplify the dynamics within the fast subspace~\cite{ForsterBook,Pomeau1975,Grabert1982}.

To reap the benefits of the dissipation, we represent $\ket{\tilde q_j(t)}$ as an MPS. The unitary part of the evolution can then be done with standard MPS techniques; for the nearest-neighbor Hamiltonians studied below, the time-evolving block decimation (TEBD) algorithm~\cite{VidalTEBD,MurgReview,Schollwock2011,Paeckel2019} provides an efficient solution. In this language, the superoperator $\op{D}_{\ell_*,\gamma}$ becomes a matrix product operator (MPO)~\cite{MurgReview,Pirvu10,Schollwock2011}. One can then straightforwardly evaluate Eq.~\eqref{eq:DissipativeEvol}, as illustrated in Fig.~\ref{fig:Method}(c). As we will show, this can be done accurately with relatively low bond dimension, even for large systems and long times, provided that the dissipation is sufficiently strong.

$\op{D}_{\ell_*,\gamma}$ in fact has an exact MPO representation with bond dimension $\ell_*+1$. We label the local basis `states' by $n = \Id, X, Y, Z$ (generalization to higher spin is straightforward). We then write the local MPO tensor, $W^{nn'}_{ab}$, as a matrix acting on the virtual indices $a,b = 0,1,\ldots,\ell_*$. They read $W^{\Id\Id}_{ab} = \delta_{a=b}$ and $W^{XX}_{ab} = W^{YY}_{ab} = W^{ZZ}_{ab} = \delta_{a=b-1} + e^{-\gamma} \delta_{a=b=\ell_*}$, all others being zero. The MPO is contracted with the vector $v_L = (1,0,\ldots,0)$ on the left, and $v_R = (1,\ldots,1,1)$ on the right. It is easy to check that this reproduces the desired result.

\begin{figure}
\centering
	\includegraphics[width=1.\columnwidth]{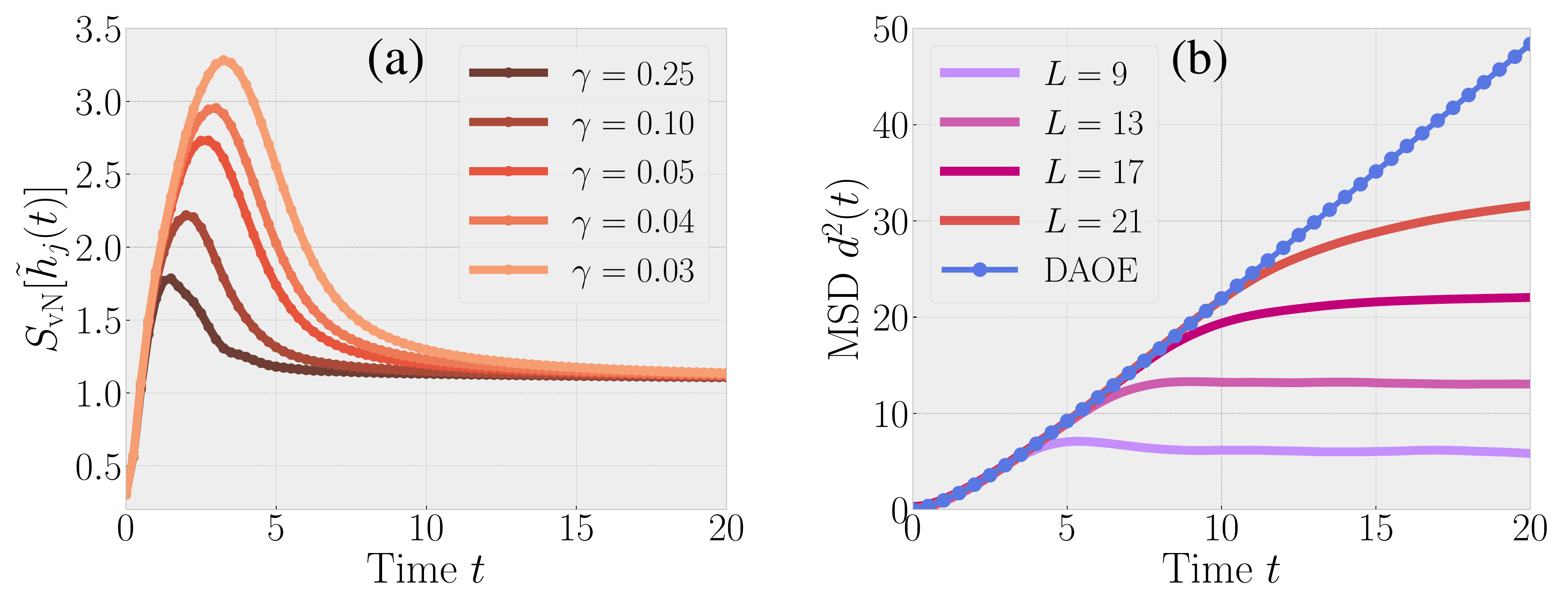}
	\caption{Testing DAOE on the Ising model~\eqref{eq:IsingDef}. (a) shows how the dissipation (for $\ell_*=2$, $\Delta t = 0.25$) suppresses operator entanglement (measured in units of $\ln{2}$). (b) shows that the MSD~\eqref{eq:MSD} is correctly captured to long times by the DAOE (same $\ell_*,\Delta t$; $\gamma = 0.03$, using bond dimensions $\chi=512$), by comparing to exact results on small chains ($L=9,13,17,21$).}
	\label{fig:Test}
\end{figure}

The main limitation in the MPS representation of $\ket{q_j}$ is the \emph{operator entanglement}~\cite{Zanardi2001,Bandyopadhyay2005,Prosen2007,Pizorn2009,Dubail2017,Zhou2017}, $S_\text{vN}[\tilde q_j(t)]$, defined as the half-chain von Neumann entropy of the normalized state $\ket{\tilde q_j(t)} / \sqrt{\braket{\tilde q_j(t) | \tilde q_j(t)}}$. For generic unitary dynamics, it tends to increase linearly~\cite{Znidaric2007,Jonay2018}, $S_\text{vN}[q_j(t)] \propto t$. In this case, the bond dimension $\chi$ needed for a faithful MPO/MPS representation grows exponentially with $t$, cutting short the times one can simulate~\cite{Schollwock2011,Paeckel2019}. We find that applying the dissipator \emph{decreases} the operator entanglement, and this effect always becomes dominant at long times (see Fig.~\ref{fig:Test}(a)). This key observation means that we can calculate $\ket{\tilde q_j(t)}$ with high precision, up to very long times, with a \emph{finite} $\chi$.

\paragraph{Results.---} We use our method to calculate the dynamical correlations between the central site $i=\frac{L+1}{2}$ (we take $L$ odd) and all other positions, $C_j(t) \equiv \tr{q_j \tilde{q}_{\frac{L+1}{2}}(t)} / \mathcal{N}$. We normalize these such that $\sum_j C_j(0) = 1$. One can characterize the spreading of correlations by the \emph{mean-square displacement} (MSD),
\begin{equation}\label{eq:MSD}
d^2(t) \equiv \sum_j C_j(t) j^2 - \left( \sum_j C_j(t) j \right)^2.
\end{equation}
In the strongly interacting, non-integrable systems we study, high-temperature transport of conserved quantities is expected to be \emph{diffusive}~\cite{Bloembergen1949,DeGennes1958,Kadanoff1963,ForsterBook}, which manifests in a linear growth of the MSD at long times, $d^2(t) \propto t$. This suggests defining a \emph{time-dependent diffusion constant}~\cite{Steinigeweg2009,Steinigeweg2017,Yan2015,Luitz2017,Bertini2020Review} as $2 D(t) \equiv \partial_t d^2(t)$. The physical diffusion constant is then $D \equiv \lim_{t\to\infty} D(t)$ (assuming $L\to\infty$ first). Further information about the frequency- and wavevector-dependence of the conductivity can be obtained by looking at space-time dependence of $C_j(t)$~\cite{Bertini2020Review,vanBeijeren1982,Delacretaz2019}.

Our approach is as follows. We calculate $D(t)$ for the dissipative evolution, and then approach the unitary dynamics by decreasing $\gamma$, while keeping $\Delta t$ and $\ell_*$ fixed. We decrease $\gamma$ until we observe signs of convergence, allowing us to extrapolate the results for $D$ back to $\gamma\to 0$. We can estimate the accuracy of this extrapolation by comparing different values of $\ell_*$. As stated above, the value of $\Delta t$ is in principle irrelevant, as one finds a scaling collapse as a function of $\gamma / \Delta t$ for small $\gamma$. However, in practice, $\Delta t$ should be small enough so that one can follow the full dynamics up to $\Delta t$ with the given bond dimension. It is also numerically more efficient not to make $\Delta t$ too small, in order to reduce the number of MPO-to-MPS multiplications we need to perform. We find that $\Delta t \approx 1$ (in units of microscopic couplings) works well. We investigate two Hamiltonians which we expect to be generic; further results on discrete circuit models are presented in \cite{suppmat}.

\paragraph{Energy transport in the Ising chain.} We first consider the Ising model in a tilted field,
\begin{align}\label{eq:IsingDef}
H = \sum_j h_j \equiv \sum_j \left( g_x X_j + g_z Z_j + \frac{Z_{j-1} Z_{j} + Z_j Z_{j+1}}{2} \right).
\end{align}
We fix $g_x = 1.4$ and $g_z = 0.9045$. At these values, we expect the model to be strongly chaotic~\cite{Karthik2007,KimHuse2013}, and hard to simulate exactly, due to fast entanglement growth. Here, $h_j$ is the energy associated to site $j$. This is the only local conserved density in the model, and its correlations capture energy (or heat) transport~\cite{KimHuse2013}. We therefore take $q_j \equiv h_j$ in this case, and evolve $h_\frac{L+1}{2}$, as an MPO, according to Eq.~\eqref{eq:DissipativeEvol}. We perform the unitary part of the dynamics with TEBD, using a small Trotter time-step $0.01$. We take large enough systems ($L=51$) such that finite size effects are negligible at the times we study.

Fig.~\ref{fig:Test}(a) confirms that the dissipation limits the operator entanglement growth, so that the entropy $S_\text{vN}[\tilde h_j(t)]$ peaks and then decreases. The time and height of the peak increase as $\gamma$ gets smaller, but for any non-zero $\gamma$, dissipation dominates at long times. Moreover, we find that after the peak, $S_\text{vN}$ approaches $1$ in units of $\ln{2}$, indicating that the operator is increasingly dominated by the local densities, $\tilde h_\frac{L+1}{2}(t) \approx \sum_j C_j(t) h_j$. 

We benchmark our method by comparing it to exact results on small systems, calculated using the \emph{canonical typicality} approach~\cite{Steinigeweg2014_1,Steinigeweg2015,Heitmann2020}, for up to $L=21$ sites. In this case, finite-size effects limit the times one can reach to $t\approx 10$. We compare these to the dissipative method for a particular set of parameters, $\ell_*=2, \Delta t = 0.25, \gamma = 0.03$, which we expect to be close to being converged to the physical diffusion constant (see below). The results for the MSD are presented in Fig.~\ref{fig:Test}(b). The curve from the dissipative evolution follows the exact results, and then continues to grow linearly to much longer times, well beyond the reach of exact numerics. This is despite the fact that at these times, the dissipation already had a large effect (as measured, for example, by the decay of $S_\text{vN}$), and $\tilde h_{\frac{L+1}{2}}(t)$ is far from the true time-evolved operator. Note that the dissipation is essential in allowing us to reach long times; for the same bond dimension ($\chi=512$), TEBD without dissipation starts deviating from the exact results around times $t \approx 7 - 8$ due to truncation errors. 

Having established the potential of the DAOE method, we now embark on the strategy outlined above, approaching the unitary limit by decreasing $\gamma$ gradually from $\gamma=\infty$. For each set of parameters, we calculate a time-dependent diffusion constant $D_{\ell_*,\Delta t}(t;\gamma)$. In the limit $\gamma\to 0$ one would recover the physical result, $\lim_{\gamma\to 0} D_{\ell_*,\Delta t}(t;\gamma) = D(t)$, for any $\ell_*$ and $\Delta t$. In practice, we are limited to some minimal $\gamma$ we can simulate with a certain bond dimension, while avoiding truncation errors. However, as we show, one can extrapolate from the data to get an estimate for the diffusion constant at $\gamma = 0$. Estimates for different $\ell_*$ then allow us to check the accuracy of this extrapolation.

\begin{figure}
\centering
	\includegraphics[width=1.\columnwidth]{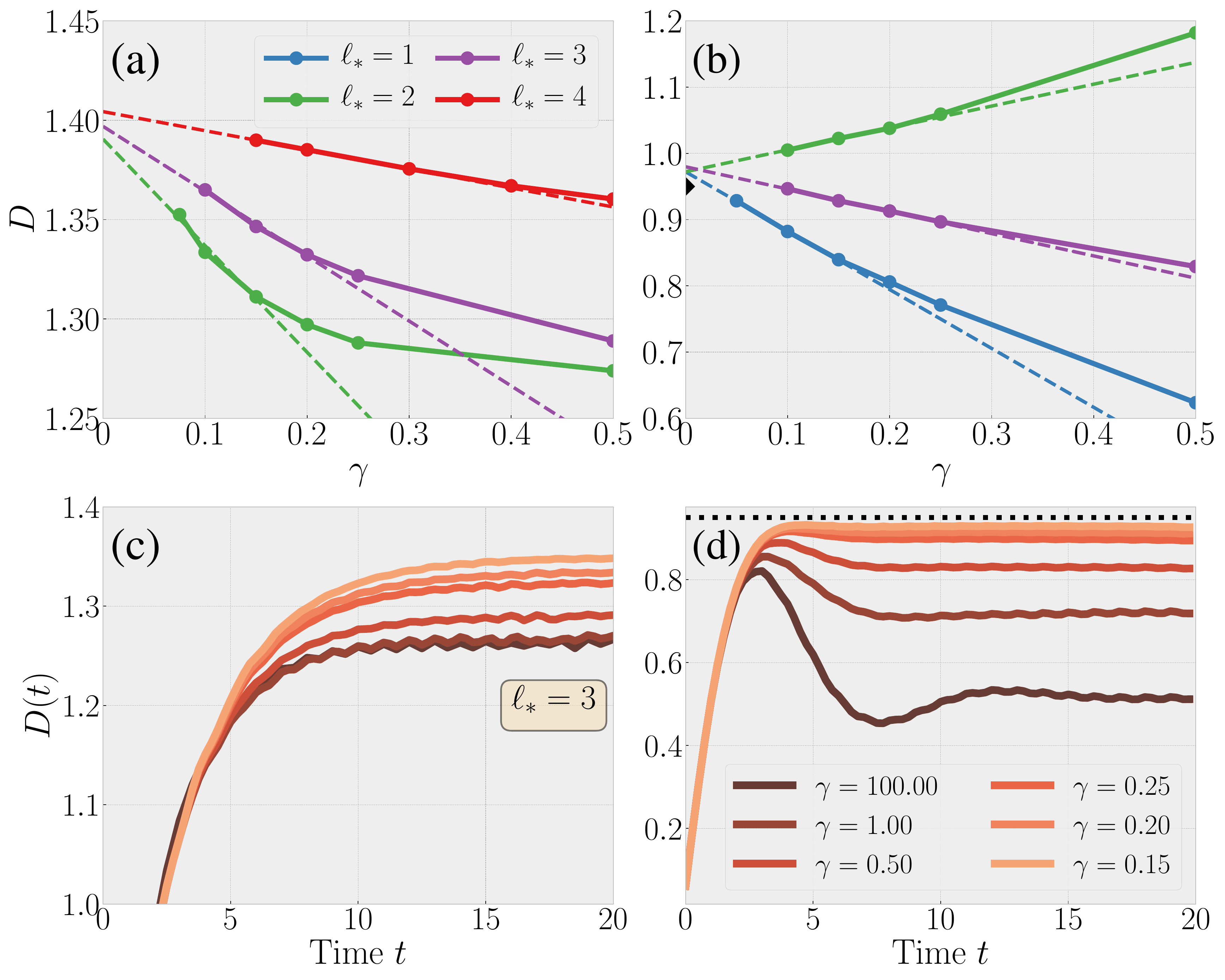}
	\caption{Estimating the diffusion constant. (a,c) show data for the Ising chain~\eqref{eq:IsingDef} and (b,d) for the XX ladder~\eqref{eq:XXladder}. We fix $\Delta t = 1$ and use bond dimensions up to $\chi=768$. In (c) and (d) we show results for the time-dependent diffusion constant at a fixed $\ell_*=3$ for varying $\gamma$, showing clear signs of convergence. In (a,b) we show the the estimate for $D$ (taken as the average of $D(t)$ in the interval $t\in [15,20]$). Data for the weakest dissipations is well fit by a linear extrapolation, and results for different $\ell_*$ give consistent estimates for the physical diffusion constant. In (b) and (d), the $\blacktriangleright$ and dotted line represent the estimate $D = 0.95$ from Ref. \onlinecite{Steinigeweg2014_2}.}
	\label{fig:Results}
\end{figure}

The results are shown in Fig.~\ref{fig:Results}(a,c), for $\Delta t = 1$ and $\ell_* = 2,3,4$. $D(t)$ saturates to a $\gamma$-dependent constant. When $\gamma$ is made sufficiently small, we find that the results converge. The last few data points are well fit by a straight line, which allows us to extrapolate $D$ back to $\gamma = 0$. The extrapolated results for different choices of $\ell_*$ all agree to within $\approx 1\%$ error, supporting our conclusions that we indeed reached the physical diffusion constant (in this case, $D \approx 1.40$). This constitutes strong evidence that our method can successfully capture transport coefficients to a high precision.

\paragraph{Spin transport in the XX ladder.} Next, we study a spin-$1/2$ model on a two-leg ladder. We denote by $j=1,\ldots, L$ the rungs of the ladder, and use $a=1,2$ for the two legs. Pauli operators on a given site are specified as $X_{j,a}$, etc. The Hamiltonian then reads
\begin{align}\label{eq:XXladder}
    H = &\sum_{j=1}^L \sum_{a=1,2}  \left( X_{j,a} X_{j+1.a} + Y_{j,a} Y_{j+1.a} \right) \nonumber \\
    +&\sum_{j=1}^L \left( X_{j,1} X_{j,2} + Y_{j,1} Y_{j,2} \right).
\end{align}
Besides energy, this model also conserves the spin $z$ component, $\sum_{j,a} Z_{j,a}$. We examine the transport of the corresponding local conserved density  $q_j = Z_j \equiv (Z_{j,1} + Z_{j,2})/2$ along the chain. We take a system of $L=41$ rungs, which is large enough to avoid finite-size effects, up to the times ($t \approx 20$) that we simulate. 

Spin transport in this model has been studied in a number of previous works, finding clear evidence of diffusive behavior with a diffusion constant $D \approx 0.95$~\cite{Steinigeweg2014_2,Karrasch2015,Kloss2018}. Here we show that our method reproduces this result on much larger systems. We perform the same analysis as in the Ising model, comparing $D$ for different $\gamma$ and extrapolating back to $\gamma=0$; the results are shown in Fig.~\ref{fig:Results}(b,d). We find that the extrapolated results are all within the range $D \approx 0.96 - 0.98$ (even for $\ell_* =1$, where energy-conservation is violated). The fact that these values are all very close to one another, and to the previous result, strongly supports the validity of our method.

\paragraph{Conclusions.---} We introduced a controlled numerical method for computing transport properties in strongly interacting quantum systems at high temperatures. Our method is based on neglecting `backflow' from complicated to simple operators. We provided a simple implementation of this method, using matrix product states, which allowed us to calculate dynamical correlations without finite-size or finite-time limitations. We demonstrated the utility of this approach on two spin models, showing that it can be used to estimate diffusion constants with high precision. An interesting open question is whether the method could be further improved by using ideas from Refs. \onlinecite{White2018} and \onlinecite{Parker2019}.

There are a variety of physical problems that would be interesting to explore with this method, such as transport in 1D quantum magnets~\cite{ljubotina17,nardis20,Dupont20,Blundell15}, disordered models~\cite{Agarwal2015,BarLev2015,Potter2015,Vosk2015,Znidaric2016} or long-range interacting~\cite{Schuckert2020} systems, where existing methods are even more limited. There might also be applications in quantum chemistry, where tensor network methods are becoming increasingly important~\cite{White1999,Marti2010,Wouters2014,Yanai2014,Szalay2015}. A natural extension of our method is to finite temperatures. We expect it to work well at high temperatures, where the thermal density matrix is dominated by short operators~\cite{Araki1969,Park1995,Wolf2008,Kliesch2014,Molnar2015,Kuwahara2019}, while it presumably breaks down as the low-temperature limit is approached. Precisely when and how this happens is itself an interesting question.

\emph{Acknowledgements.}--The authors thank David Huse, Philipp Dumitrescu, Tarun Grover, Andrew Green, Fabian Heidrich-Meisner, Sean Hartnoll, Vedika Khemani, Mingee Chung, Xiangyu Cao and Daniel Parker for stimulating discussions, and in particular Ehud Altman for his \href{http://online.kitp.ucsb.edu/online/dynq_c18/altman/}{talk} at the KITP Conference ``Novel Approaches to Quantum Dynamics'' that in part inspired our work.
CvK is supported by a Birmingham Fellowship. FP is funded by the European Research Council (ERC) under the European Unions Horizon 2020 research and innovation program (grant agreement No. 771537). FP acknowledges the support of the DFG Research Unit FOR 1807 through grants no. PO 1370/2-1, TRR80, and the Deutsche Forschungsgemeinschaft (DFG, German Research Foundation) under Germany's Excellence Strategy EXC-2111-390814868. This work was initiated at KITP where TR, CvK, and FP were supported in part by the National Science Foundation under Grant No. NSF PHY-1748958 (KITP) during the ``Dynamics of Quantum Information'' program. TR further acknowledges the hospitality of KITP as part of the graduate fellowship program of the Fall of 2019, during which some of this work was performed. 

\bibliography{main}

\newpage
\onecolumngrid

\renewcommand{\thesection}{\Roman{section}}
\renewcommand{\thesubsection}{\Alph{subsection}}

\begin{center}
{\large \textbf{Supplementary Material for ``\inserttitle''\\}}
\end{center}

\section{Additional data for the Ising chain and XX ladder models}

\subsection{Convergence with bond dimension}

In the main text, we showed that the dissipation leads to a decay of the operator entanglement at long times. Crucially, this makes the maximal operator entanglement encountered during the evolution independent of system size, depending only on the parameters of the dissipation. As we argued, we can therefore capture the diffusive spreading of correlations up to arbitrarily long times, without significant finite-size or truncation effects. Here we show explicitly how the curves for $D(t)$ converge as we increase the bond dimension $\chi$. 

\begin{figure}
\centering
	\includegraphics[width=1.\columnwidth]{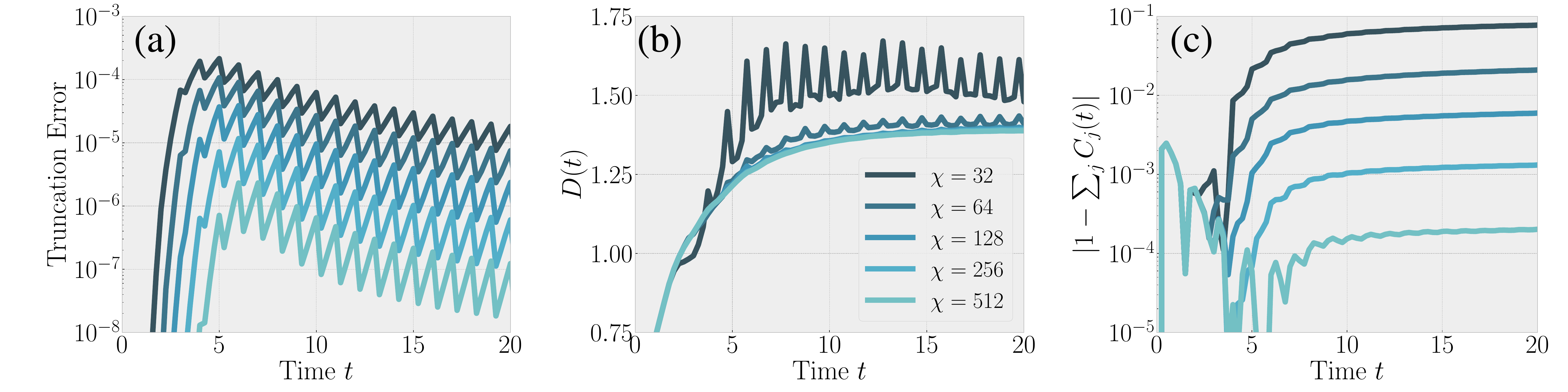}
	\caption{Convergence of results with bond dimension $\chi$ in the Ising chain~\eqref{eq:IsingDef} for dissipation parameters $\ell_* = 4$, $\Delta t =1$, $\gamma = 0.2$. (a): Truncation error per TEBD step, summed over all bonds in the chain ($L=51$ sites). (b) Convergence of results for $D(t)$ (see main text for definition). (c) Errors in the energy conservation, as measured by the sum of the coefficients of local energy density terms $C_j(t)$. }
	\label{fig:ChiConv}
\end{figure}

The results are shown in Fig.~\ref{fig:ChiConv} for the tilted field Ising model. We fix parameters $\ell_* = 4$, $\Delta t =1$, $\gamma = 0.2$ (same as in Fig.~\ref{fig:Test}(b)) and compare results for different bond dimensions $\chi=32,64,128,256,512$. As the operator entanglement peaks and decreases (see Fig.~\ref{fig:Test}(a)), the truncation error of the unitary TEBD time step also starts decreasing. While for small $\chi$, the truncation errors encountered around the peak time are already significant, they decrease (roughly linearly) with $\chi$. This also shows up in the results for the time-dependent diffusion constant, $D(t)$. While at small $\chi$ the truncation effects are clearly visible, the curves quickly converge as $\chi$ is increased. 

Another way of testing the effects of truncation is by looking at whether the conservation law (in this case, of energy) is satisfied. We consider the correlations $C_j(t)$ and normalize them such that $\sum_j C_j(0)=1$. The exact dissipative dynamics would maintain this normalization at all subsequent times due to energy conservation (assuming $\ell_*$ is larger than the support of the terms in the Hamiltonian, in this case $\ell_* \geq 2)$. This is crucial for correctly capturing transport properties. We find that the errors in the conservation law, as measured by $\left| 1 - \sum_j C_j(t) \right|$ quickly decrease as $\chi$ becomes larger. We conclude that it is possible to simulate the dissipative dynamics~\eqref{eq:DissipativeEvol} up to long times, with a bond dimension that is independent of total system size. 

\subsection{Scaling collapse as a function of $\gamma / \Delta t$}

Here, we justify our claim in the main text that when $\gamma$ is sufficiently small, the results (in particular, estimates of $D$) are functions of the ratio $\gamma / \Delta t$ only. This can be seen by utilizing the Baker-Campbell-Hausdorff formula to rewrite the evolution operator~\eqref{eq:DissipativeEvol} as 
\begin{align}\label{eq:BCH}
\left( \op{D}_{\ell_*,\gamma}  e^{i\mathcal{L} \Delta t}\right)^N \equiv \left( e^{-\op{K}_{\ell_*}\gamma}  e^{i\mathcal{L} \Delta t}\right)^N = \left( e^{-\op{K}_{\ell_*}\gamma + i\mathcal{L} \Delta t + O(\gamma \Delta t)} \right)^N = e^{-\op{K}_{\ell_*} N \gamma + i\mathcal{L} N \Delta t + O(\gamma N \Delta t)} = e^{t \left( i \op{L} - \op{K}_{\ell_*} \frac{\gamma}{\Delta t} \right) + O(\gamma t)}, 
\end{align}
where $t = N\Delta t$ and we have introduced the logarithm of the dissipator, acting on a Pauli string as
\begin{align}
\op{K}_{\ell_*}[\op{S}] = 
\begin{cases}
0 &\text{ if } \ell_\op{S} \leq \ell_* \\ 
(\ell_\op{S} - \ell_*) \op{S} &\text{ otherwise}.
\end{cases}
\end{align}
In the second equality of Eq.~\eqref{eq:BCH} we assumed $\gamma \ll 1$ to drop higher order terms that scale as $\gamma^2 \Delta t$. We also assume that $\Delta t$ is at most an $O(1)$ quantity, so that terms that scale as $\gamma \Delta t^2$ are of the same order as $\gamma \Delta t$. 

Eq.~\eqref{eq:BCH} shows that the dynamics only depends on the ratio $\gamma/\Delta t$, and not on the individual value of $\gamma$ and $\Delta t$, up to times $t \approx 1/\gamma$. As such, it does not directly constrain the diffusion constant, which is extracted from the long-time limit. However, in practice we find that $D(t)$ saturates to a constant at a finite time $t_\text{sat}$. While $t_\text{sat}$ itself depends on $\gamma$ and $\Delta t$ (as well as on the Hamiltonian), we find that this dependence is relatively weak; in particular, $t_\text{sat}$ should converge to a finite, $O(1)$ value as $\gamma\to 0$. Therefore, estimate of $D$ should also depend only on the ratio $\gamma/\Delta t$, provided that we are in the regime where $\gamma t_\text{sat} \lesssim 1$.

\begin{figure}
\centering
	\includegraphics[width=1.\columnwidth]{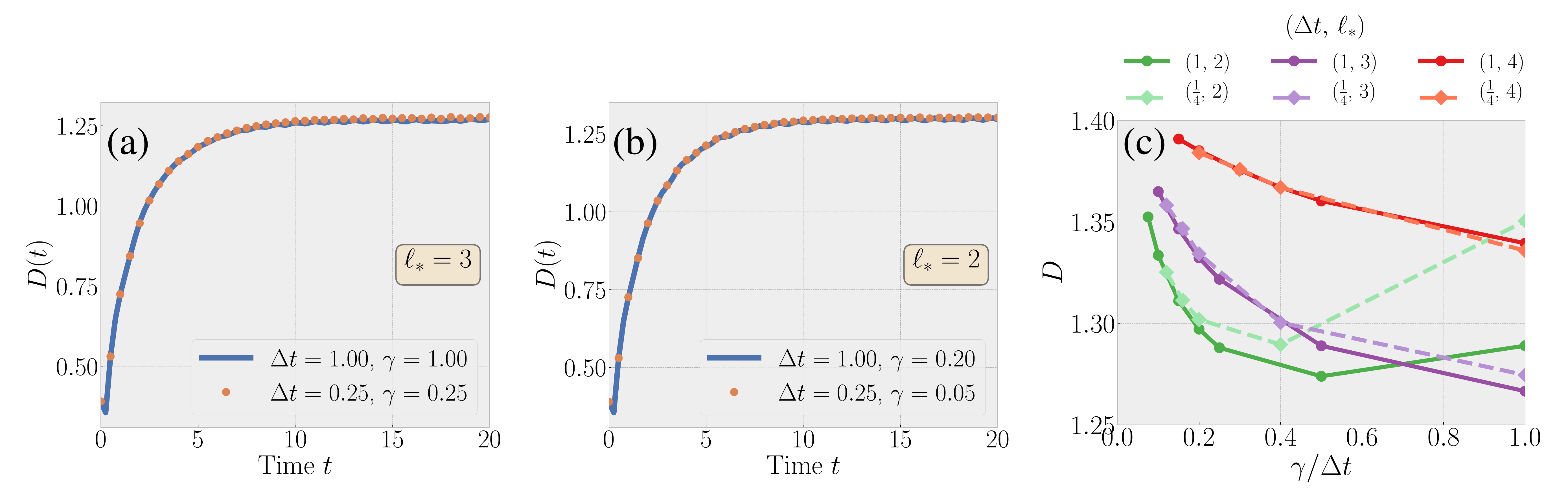}
	\caption{Scaling collapse as a function of $\gamma / \Delta t$. (a,b): comparison of time-dependent diffusion constants for two curves with different $\Delta t$ but the same ratio $\gamma / \Delta t$. When $\gamma$ is sufficiently small, the results remain close to each other even at long times. (c): Estimates of $D \equiv 
	lim_{t\to\infty} D(t)$, comparing $\Delta t = 1$ and $\Delta t = 1/4$. In the small $\gamma$ regime, relevant for extrapolation, the curves with the same $\ell_*$ collapse when plotted as function of $\gamma / \Delta t$.}
	\label{fig:scaling}
\end{figure}

Testing this expectation on the Ising chain~\eqref{eq:IsingDef}, we find that it works remarkably well, even for $\gamma \approx 1$ (we also find that it works increasingly well as $\ell_*$ gets larger).  This is shown in Fig.~\ref{fig:scaling}. Figs.~\ref{fig:scaling}(a,b) show that curves with identical ratio $\gamma/\Delta t$ are the same at early times, and, moreover, their late time saturation values are also close to one another, provided that we are in a regime with sufficiently small $\gamma$. Consequently, the estimates for $D$ show a scaling collapse when data for the same $\ell_*$ but different $\Delta t$, are plotted as a function of $\gamma / \Delta t$, see Fig.~\ref{fig:scaling}(c).

\subsection{Operator weights}

In the main text, we noted that the operator von Neumann entropy of the dissipatively evolving local density approaches $1$ (in units of $\ln{2}$) at long times. Our interpretation was that this points to a long-time behavior where the evolving operator is increasingly dominated by its diffusive, `conserved' part, $\tilde q_0(t) \approx \sum_j C_j(t) q_j$. We now further support this by calculating the weight of various operators in $\tilde q_0$ (in this section we use a different notation from the main text, with $0$ denoting the center site).

To define what we mean by the weight of an operator, let us expand $\tilde q_0$ in the basis of Pauli strings, $\tilde q_0 = \sum_{\op{S}} c_\op{S}(t) \op{S}$; the weight of the Pauli string is then the squared coefficient, $|c_\op{S}|^2$. The total weight on operators with length $\ell$ is given by the following quantity:
\begin{equation}
    W_\ell(t) \equiv \sum_{\substack{\op{S} \\ \ell_\op{S} = \ell}} |c_\op{S}(t)|^2.
\end{equation}
For unitary evolution one would have a conserved total weight, $\sum_\op{S} |c_\op{S}(t)|^2 = \sum_\ell W_\ell(t) = 1$. During evolution, the weight gets redistributed from short operators to an essentially random superposition of long ones, such that at time $t$ the operator is dominated by strings of length $\ell \sim v_\text{B} t$, with $v_\text{B}$ the butterfly velocity. This leads to the linear growth of operator entanglement with time.

The dissipator fundamentally changes this picture, as it \emph{removes} operator weight from long strings. This reverses the effect of the unitary dynamics, making the contribution of short operators dominant at long times, which leads to the observed decay in the entanglement. While short operators, with $\ell \leq \ell_*$, are not affected directly by the dissipator, their weight also decreases as they get converted into longer strings which are subsequently dissipated. However, due to the hydrodynamic nature of transport, we find that the weight associated to local densities, $|C_j|^2 \equiv |c_{q_j}|^2$ decreases parametrically more slowly than those of non-conserved operators, so that they dominate at long times. 

To show this, we consider the XX ladder~\eqref{eq:XXladder} and consider the evolution of the spin density, $\tilde Z_0(t)$. Calculating operator weights for this object, we find that the weight on local densities decays as $W_{\ell=1} \sim t^{-1/2}$, as expected from the diffusive nature of spin transport~\cite{RvK2018,Khemani2018}. Considering larger $\ell$, we find two things. First, for $\ell > \ell_*$, the weight decreases exponentially with $\ell$, as expected from the form of the dissipator. More importantly, however, we find that the weights for $\ell > 1$ decay parametrically faster in time, $W_{\ell>1} \sim t^{-3/2}$ (even when $1 < \ell \leq \ell_*$); this is shown in Fig.~{}. This is consistent with our earlier prediction, that $\tilde Z_0(t)$ is dominated by the local densities at long times. 

\begin{figure}
	\centering
	\includegraphics[width=0.7\textwidth]{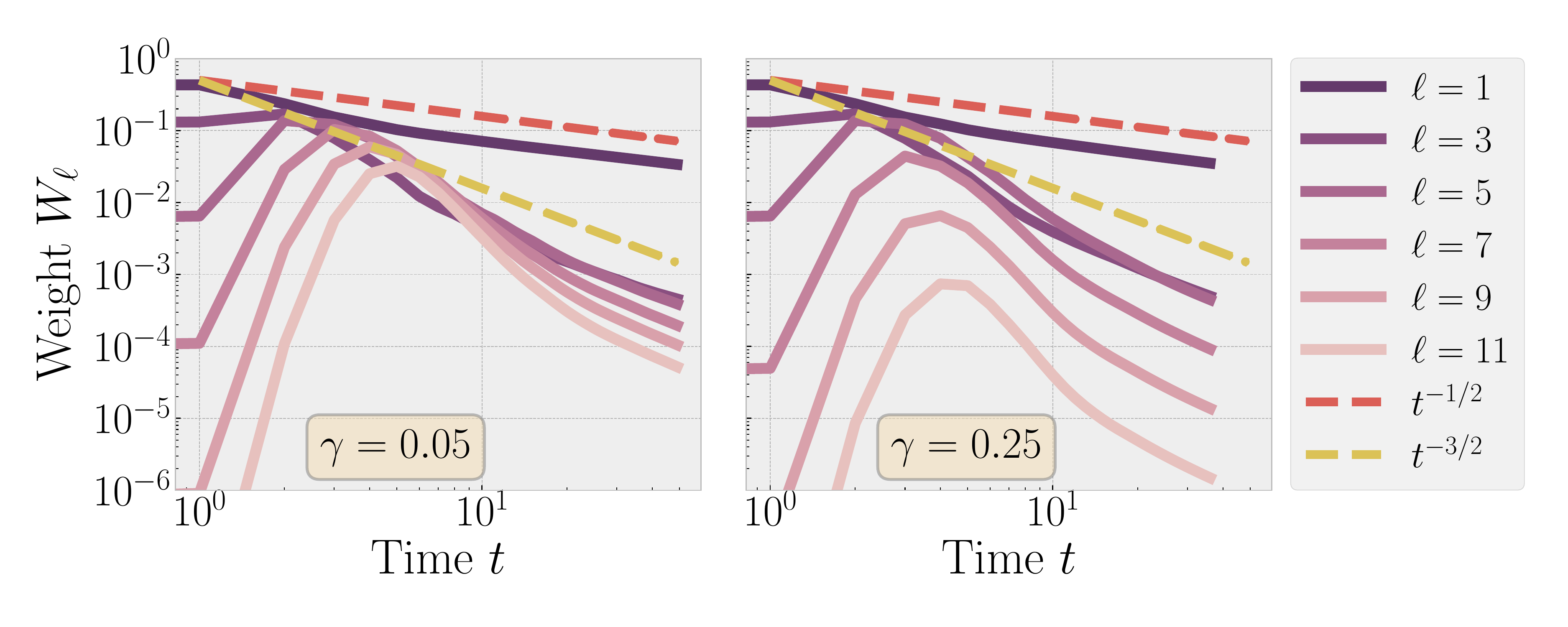}
	\caption{Total weight on strings of size $\ell$ as a function of time. The majority of the remaining (not yet dissipated) weight is on 1-site strings as decays as $t^{-1/2}$. The weight of longer strings decays as $t^{-3/2}$. Data shown for $\Delta t = 1$, $\ell_* = 5$ with $\gamma = 0.05$ (left) and $\gamma = 0.25$ (right).}
	\label{fig:weights}
\end{figure}

The $t^{-3/2}$ power law can be explained using the operator spreading picture developed in Refs. \onlinecite{RvK2018,Khemani2018}. In this picture, we rewrite the time-evolved local density at position $x$ as $q_x(t)$
\begin{equation*}
q_{x}(t)=q_{x}^{\text{D}}(t)+q_{x}^{\text{B}}(t)    
\end{equation*}
where $q_{x}^{\text{D}}(t)\equiv\sum_{y}C(x,y,t) Z_{y}$ is the diffusive part of the operator and we assume that $C(x,y,t) \equiv \braket{q_y|q_x(t)}$ is well approximated by an unbiased diffusion kernel. $q_{x}^\text{B}(t)$ contains the contributions from all remaining Pauli strings, and is dominated by those with length $\ell = 2 v_\text{B}t$, with $v_\text{B}$ the operator butterfly velocity~\cite{RvK2017,Nahum2017}. The unitary dynamics leads to a conversion of weight from the diffusive to the ballistic part, whose local rate is given by `current' squared, $|\partial_y C(x,y,t)|^2$. In this way, at each time step $q_x^\text{D}$ sources new ballistically growing operators which thereafter form part of $q_{x}^{\text{B}}$. This picture can be used to deduce the behaviour of $W_\ell$ as a function of time. According to the above picture, operators of support $\ell$ would correspond to terms in $q^{\text{B}}$ which have been ballistically growing for a time interval $t-\tau=\ell/(2v_{\text{B}})$. The weight of such terms is therefore expected to be
\begin{equation*}
\int \text{d}y\left(\partial_{y}C(x,y,\tau)\right)^{2}	\sim\left[D\left(t-\frac{\ell}{2v_{\text{B}}}\right)\right]^{-3/2}.    
\end{equation*}
This shows that the weight on length $\ell$ operators at time $t\gg\frac{\ell}{2v_{\text{B}}} $ goes as $(Dt)^{-3/2}$.

\section{Spin diffusion in Floquet circuits}

\begin{figure}
\centering
	\includegraphics[width=1.\columnwidth]{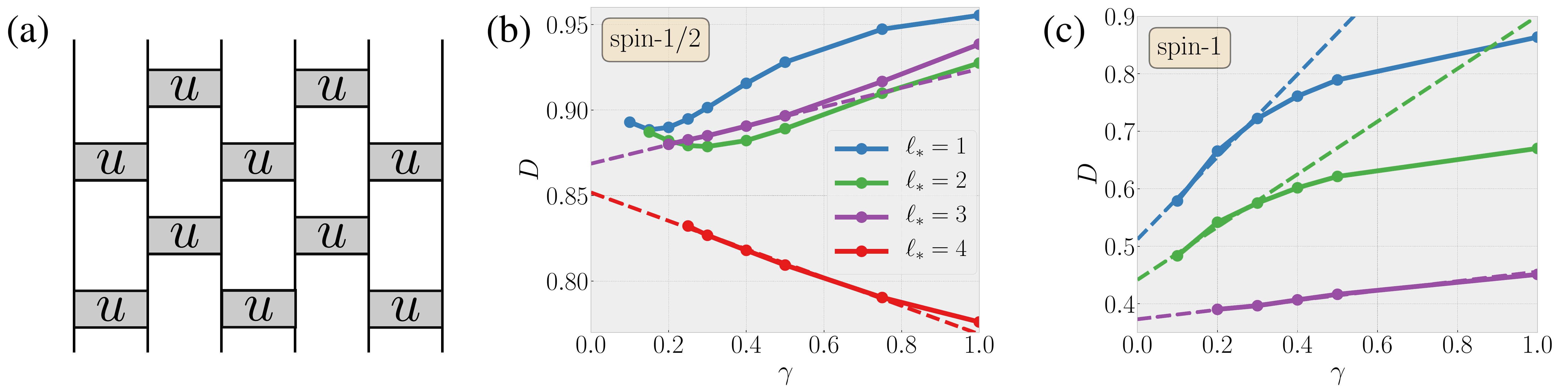}
	\caption{Diffusion constants in Floquet curcits. (a) The circuits have a brick-wall structure, updating even/odd bonds in turn. Every gate is given by the same $S_z$-conserving two-site unitary $u$. (b,c) Estimates of the spin diffusion constant for the circuit defined by Eq.~\eqref{eq:CircuitDef}, for spin-$1/2$ and spin-$1$ chains.} 
	\label{fig:circuits}
\end{figure}

We now complement the results shown for energy-conserving, Hamiltonian dynamics in the main text, with data on time-periodic models. We construct these as circuits of local unitary gates, with a `brick-wall' structure and consequently, a strict light cone. This structure is illustrated in Fig.~\ref{fig:circuits}(a). We use the same two-site unitary $u$ in each gate, such that the system has translation invariance in space (with unit cells composed of two sites) and in time (by two layers of the circuit).

We want our circuit to conserve the total spin-$z$ component. For a spin-$1/2$ chain, such a circuit is fully parametrized by three numbers, and it corresponds to a Trotterized version of an XXZ chain with a staggered magnetic field
\begin{equation}\label{eq:CircuitDef}
    u = e^{-i\left( J_{xy} (S^x_1S^x_2+S^y_1S^y_2) + J_{zz}S^z_1S^z_2 + g(S^z_1-S^z_2)\right)},
\end{equation}
where we have now used spin operator $S^\alpha$ instead of Pauli matrices (the two differ by a factor of $2$), and the subscripts refer to the two sites on which the gate acts. We choose irrational values of the three couplings, $J_{xy} = 2\sqrt{7}$, $J_{zz} = 2\sqrt{5}$, $g=2\sqrt{3}$.

We apply our dissipative evolution method for this circuit model, applying the dissipator after every second layer of the circuit (i.e., one Floquet period). We extract spin diffusion constant in the same way as in the main text. The results for the spin-$1/2$ circuit are plotted in Fig.~\ref{fig:circuits}(b). We find that the convergence to $\gamma=0$ is less clear than in the Hamiltonian models we studied in the main text. In particular, for $\ell_* =1,2$ we observe a strong non-monotonicity with $\gamma$, while $\ell_*=3,4$ do appear to converge linearly to compatible values of $D$. Nevertheless, we note that the variations in $D$ are all relatively small. 

Our interpretation is that the apparent lack of convergence in Fig.~\ref{fig:circuits}(b) is not related to the Floquet circuit nature of our model; rather, it has to do with the fact that it is close to an integrable point. It was recently shown that for $g=0$, the model in Eq.~\eqref{eq:CircuitDef} is integrable; this is closely related to the integrability of the XXZ Hamiltonian. In the latter case, a staggered field is known to break integrability~\cite{Huang2013,Steinigeweg2015,Mendoza2015}, so we expect that for generic $g$ our circuit is also non-integrable. However, we believe that the nearby integrable point is responsible for the non-trivial behavior we observe (for example some almost-conserved operator of length $\ell=3$ could explain why the $\ell_* \leq 2$ curves  have a qualitatively different behavior from $\ell_* \geq 3$).

To test this intuition, we also consider the spin-$1$ version of the same model. That is, we use the same definition of the two-site gate as in Eq.~\eqref{eq:CircuitDef}, but with $S_{1,2}^\alpha$ standing for spin-$1$ operators. The results for this case are shown in Fig.~\ref{fig:circuits}(c). While we find that getting to smaller $\gamma$ becomes quite difficult in this case, due to a quick initial growth of operator entanglement, so that our results are not as precisely converged as for the models presented in the main text, we find no evidence of strong non-monotonicities in the regime we can simulate. This reinforces our belief that the peculiar behavior exhibited by the spin-$1/2$ model is tied to the presence of nearby integrable points.

\end{document}